# A PROTOTYPE OF THE UAL 2.0 APPLICATION TOOLKIT

N.Malitsky, J.Smith, J.Wei, BNL, Upton, NY 11973, USA

Abstract

The paper presents a prototype of the accelerator commissioning and simulation application toolkit based on the Unified Accelerator Libraries (UAL) framework.

## 1 RATIONALE

The initial version of the Unified Accelerator Libraries [1] has been developed to address the realistic beam dynamics study including an unlimited combination of physical effects and dynamic processes. This goal has been achieved by introducing an open infrastructure where diverse accelerator approaches and programs are implemented as separate C++ libraries or Perl modules connected together via Common Accelerator Objects (such as Element, Twiss, Particle, *etc*.). A universal homogeneous shell based on the Perl extension mechanism allows scientists to integrate and manage all these C++ and Perl components in the single development and research environment. At this time, the UAL 1.x off-line simulation environment joins six object-oriented accelerator programs and is being developed for modeling of the low beam loss in the Spallation Neutron Source (SNS) Ring [2-4].

The new UAL 2.0 application toolkit aims to apply the proven UAL 1.x analysis and design patterns to commissioning tasks. In comparison with the off-line simulation development, the implementation of commissioning applications is associated with additional technical issues, such as the client-server infrastructure, Graphical User Interface (GUI), database connectivity, and others. Until recently, developers had to deal with a variety of vendor-specific products or proprietary solutions. Java 2 offers a standard-based consolidated platform covering almost all necessary components excluding accelerator-specific interfaces and algorithms. In this paper, we present how Java technologies can be leveraged for integrating the UAL application framework into the accelerator control system environment.

## 2 ARCHITECTURE

In the UAL approach, the accelerator application environment has been divided into three functional layers: high level applications, middle layer, and data resources (see Fig. 1):

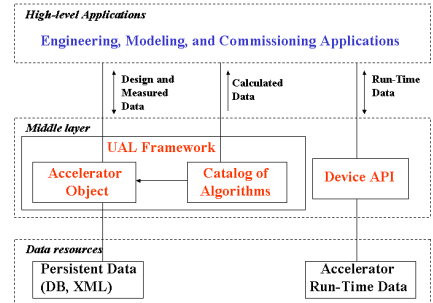

Figure 1: The functional diagram of the accelerator application environment.

The middle layer forms the basis of the application infrastructure. It aims to provide high-level applications with a uniform interface to heterogeneous distributed data resources. According to our previous work [5], all accelerator data are grouped into three categories:
- *design and measured static accelerator parameters*, such as locations of elements, magnetic fields, *etc*.;
- *calculated data,* such as transfer matrices, *etc*.;
- *device run-time data*, such as detector signals, *etc*.

For each category, the middle layer supplies the corresponding module: accelerator object, catalog of algorithms, and device API. This modular Accelerator-Algorithm-Device structure is driven by the use-case analysis of existing accelerator applications. We consider three major types of accelerator software:
- *engineering tools* (e.g. MEDM) dealing with some variant of the device API module for direct control and monitoring of device run-time data;
- *off-line modeling programs* (e.g. UAL 1.x, MAD) that are traditionally built from a combination of accelerator model containers and a set of associated analysis and tracking algorithms.
- *commissioning applications* (e.g. Orbit Difference Display) that can be considered as a complex mixture of components from two previous categories.

Building different tools from common modules preserves integrity of whole application environment. On the other hand, the modular structure of the middle layer permits its partial integration with existing control facilities and frameworks.

The middle layer is usually associated with the application server. In the prototype version of the UAL 2.0 toolkit, it has been developed as a set of Java local

packages that can be deployed in different modules of application systems (see Fig.2):

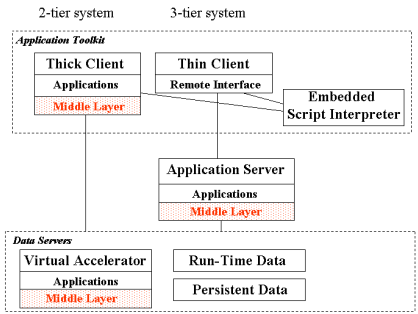

Figure 2: The deployment diagram of the accelerator application environment.

The application toolkit is designed after the combined 2-tier/3-tier scheme using the common Swing-based graphical user interface. This approach has been chosen to provide the most comprehensive and extensible environment for developing and integrating various accelerator applications that can be implemented as:
- *Java thick local modules* (e.g. analysis, graphics) running in the same process and sharing common objects;
- *Java thin clients* connected to remote application servers;
- *embedded scripts* (e.g. Jython, JavaScript) for rapid prototyping and extensions.

The appearance of the Web-based deployment tools (such as Sun Java Web Start, and others) facilitates the maintenance and distribution of the Swing desktop applications through the Web network and makes their mobility practically comparable with traditional Web solutions.

## 3 MIDDLE LAYER

The middle layer is formed from the connection of the UAL Element-Algorithm-Probe framework [6] with the Device API of the control systems. The Element-Algorithm-Probe scheme is based on the fundamental description of the physical experiment representing physical process as interactions between Elements and Probes. In the UAL applications, Element is associated with a *persistent* identified entity such as a magnet or RF cavity. Probe is a *serialized* container that can represent a bunch of particles or some mathematical abstraction (e.g. Twiss parameters, Taylor map) exchanged and transformed by a set of Algorithms. Element and Algorithm create a basis of two middle layer modules: Accelerator Object and Catalog of Algorithms.

### 3.1 Accelerator Object

The accelerator object is implemented after the Standard Machine Format (SMF[7]) model. Its structure has produced as a result of analysis and generalization of several accelerator formats. SMF preserves all Standard Input Format element types and provides a uniform approach for accommodating new elements and attribute sets. The SMF object model has also been optimized from the implementation perspective. Its data structures and associations are described via standard containers and design patterns. This description allowed us to apply the proven mapping techniques for implementing the SMF objects into various physical representations, such as Java classes, XML[7], relational database tables [8], and GUI components. Fig. 3 shows the Accelerator Manager, one of the common modules of the UAL 2.0 application toolkit.

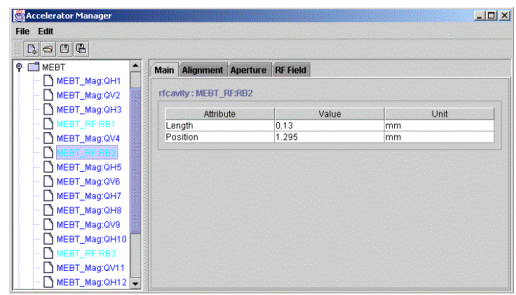

Figure 3: The Accelerator Manager.

Accelerator Manager is the editor and browser of accelerator object data. Also it can be considered as a visual illustration of the SMF model and its mapping into the Swing components. The view of the accelerator hierarchical structure is displayed and navigated by the JTree class. The accelerator node data are represented by an open collection of the JTabbedPane's panels where each tab panel is associated with a single attribute set (e.g. magnetic field, aperture, alignment, etc.). Different attribute sets have the same structure, an associative array of simple and indexed properties, and can be accessed via the generic interface of the JTable component.

A uniform organization of accelerator nodes and attribute sets has been leveraged for developing other generic applications, for example, connectors between the accelerator containers and data sources. However, the generic interface may not be sufficient enough for simulation and modeling tasks. To address requirements of this category of applications, each attribute set has been augmented with the specialized direct interface to its data members.

## 3.2 Catalog of Algorithms

The connection of Algorithms with Accelerator Object is implemented after the Strategy and Registry (Finder) patterns. According to the Strategy pattern algorithms are encapsulated in separate classes associated with the corresponding element type. A collection of these associations is organized in Algorithm Registry. There could be many different algorithm registries depending on the application type, selected approach, accelerator structure, and other factors. Despite different implementations, all of them share the same interface forming an open catalog of accelerator algorithms.

## 3.3 Device API

Accelerator Device represents an identified accelerator physical entity which run-time parameters can be accessed, controlled, and monitored by accelerator applications. There is a variety of different site-specific device descriptions ranging from generic ("narrow") to explicit ("wide") interfaces. The UAL toolkit is being developed for the SNS project which control system is based on the EPICS infrastructure. The interface to the EPICS run-time database is provided by the Channel Access (CA) library implemented in the C language. In addition to the CA original version, the EPICS software includes its Java wrapper, JCA [9], based on the Java Native Interface technology. JCA offers most of the synchronous and asynchronous CA features and has been integrated in the middle layer of the UAL 2.0 application toolkit.

# 3 GRAPHICAL USER INTERFACE

The GUI framework of the UAL toolkit is designed after the NSLS Application Manager approach [10] that provides the uniform navigation interface to commissioning and engineering applications (see Fig. 4). The main window of the Application Manager is based on the popular Windows Explorer layout. The left side displays the hierarchical structure of application categories, and the right panel provides a list of table rows with buttons for starting applications of selected category. In additional to the traditional Explorer functionality, the Application Manager allows user to combine together the different types of accelerator applications: external programs (e.g. MEDM engineering screen), local modules (e.g. Accelerator Manager), embedded scripts (e.g. Java Python), and others. The structure of the project applications is flexible and specified in the external XML file.

# 4 ACKNOLEDGMENTS

The authors would like to thank the SNS Accelerator Physics and Control groups for many useful discussions and suggestions.

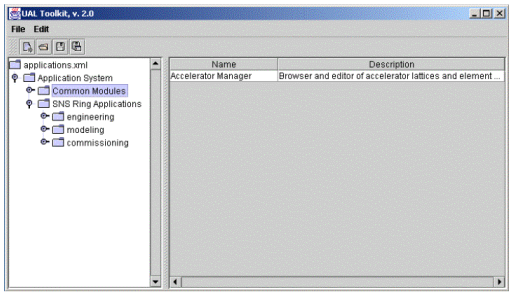

Figure 4: The Application Manager